# High-Mg Calcite Nanoparticles Within a Low-Mg Calcite Matrix via Spinodal Decomposition: A Widespread Phenomenon in Biomineralization

Nuphar Bianco-Stein[a], Iryna Polishchuk[a], Arad Lang[a], Lotan Portal[a], Catherine Dejoie[b] and Boaz Pokroy[a]*

[a] Department of Materials Science and Engineering and the Russell Berrie Nanotechnology Institute, Technion−Israel Institute of Technology, 32000 Haifa, Israel.

[b] ESRF-The European Synchrotron, CS 40220, 38043 Grenoble Cedex 9, France.

## Abstract

During the process of biomineralization, organisms utilize various biostrategies to enhance the mechanical durability of their skeletons. In this work, we establish that the presence of high-Mg nanoparticles embedded within lower Mg-calcite matrices is a widespread strategy utilized by various organisms from different kingdoms and phyla to improve the mechanical properties of their high-Mg calcite skeletons. We show that such phase separation and the formation of high-Mg nanoparticles are achieved through spinodal decomposition of an amorphous Mg-calcite precursor. Such decomposition is independent of the biological characteristics of the studied organisms belonging to different phyla and even kingdoms, but rather originates from their similar chemical composition and a specific Mg content within their skeletons, which generally ranges from 14 to 48 mol% of Mg. We show evidence of high-Mg calcite nanoparticles in the cases of 6 biologically different organisms all demonstrating more than 14 mol% Mg-calcite, and consider it *likely* that this phenomenon is immeasurably more prevalent in nature. We also establish the absence of these high-Mg nanoparticles in organisms whose Mg content is lower than 14 mol%, providing further evidence that whether or not spinodal decomposition of an amorphous Mg-calcite precursor takes place is determined by the amount of Mg it contains. The valuable knowledge gained from this biostrategy significantly impacts the understanding of how biominerals, though comprised of intrinsically brittle materials, can effectively resist fracture.

## Significance Statement

Biominerals are extraordinarily intricate and possess superior mechanical properties as compared to their synthetic counterparts. In this study, we show that the presence of high-Mg calcite nanoparticles within a low-Mg calcite matrix is a widespread phenomenon among marine organisms whose skeletons are comprised of high-Mg calcite. It seems most likely that formation of such a complex structure is possible because of the phase separation that occurs as a result of spinodal decomposition of an amorphous Mg-calcium carbonate precursor and is followed by crystallization. We demonstrate that the basis of such phase separation stems from chemical composition rather than from biological similarities. The presence of high-Mg calcite nanoparticles increases the skeletons' toughness, rendering them resistant to fracture.

## Main Text

Introduction

A range of organisms utilize calcite to build their skeletons while incorporating Mg into the $CaCO_3$ crystal lattice (1–5). The content of Mg within the calcite crystal varies (6), from as low as only a few mol% to levels as high as 45 mol% (7, 8). High-Mg calcite biominerals with a Mg content that exceeds the thermodynamic solubility limit are secreted by a wide variety of organisms belonging to different kingdoms and phyla. For example, high-Mg calcite is found in the skeletons of brittle stars (9), sea urchins (10), starfish (11), sea sponges (12) and corals (13), as well as in plants such as the coralline red algae (14, 15). The Mg plays a significant role in the crystallization process (16–18) and the incorporation of large amounts of Mg is facilitated by crystallization via an amorphous calcium carbonate (ACC) precursor (19–23). The incorporated Mg ions modulate the crystal properties of the $CaCO_3$, enhancing its structural performance. Owing to the smaller radii of Mg ions than that of Ca ions (24), compressive stresses are formed within the $CaCO_3$ crystal and lead to improved hardness (25, 26).

The brittle star *Ophiomastix wendtii* (9, 27) and 2 species of the coralline red algae, *Jania* sp. and *Corallina* sp. (28), were the focus of our recent studies on high-Mg calcite biomineralization. The brittle star *O. wendtii* (phylum Echinodermata) is an echinoderm that is highly sensitive to light and possesses a remarkable visualization system composed of micrometer-sized Mg-calcite lenses located on its dorsal arm plates, all directed with their c-axis upright to avoid birefringence (29, 30). *Jania* sp. and *Corallina* sp. (phylum Rhodophyta) are highly abundant coralline red algae growing in oceans worldwide (31). When residing in shallow waters they are constantly exposed to external stresses imposed by the sea waves (32, 33). These algae incorporate hollow helical microstructures that highly increase their compliance, allowing them to better adapt to their natural environment (34). They are heavily mineralized, with Mg-calcite nanocrystals deposited on their cell walls (35–37). Despite their biological distinction, these 3 organisms demonstrate common structural aspects, among which is the presence of high-Mg nanoparticles embedded within the lower Mg-calcite matrices of their skeletons. An earlier study of *O. wendtii* by our group revealed that the high-Mg nanoparticles are coherently aligned within the lenses' low-Mg calcite matrix and are distributed in alternating layers of varying densities (9). We also observed similar structural organization in other skeletal parts of *O. wendtii*, specifically its spicules, arm vertebrae and teeth (27). In the case of both coralline red algae species, we found that the high-Mg nanoparticles were located within the poor-Mg hosting nanocrystal matrices and, in contrast to the brittle star *O. wendtii,* that they were included non-coherently (28). Additionally to our results, previous reports on the red coral *Corallium rubrum* also described Mg-calcite nanodomains within its skeleton, but they were not specified as high-Mg calcite nanoparticles (38).

Based on calculations, we proposed (27) that the formation of high-Mg nanoparticles in *O. wendtii* originates from the spinodal decomposition of a Mg-ACC precursor, and showed that such decomposition takes place in the non-stable region with a Mg content in the range of 14 to 48 mol%. This model was subsequently used by our group to explain the mechanism of formation of the high-Mg nanoparticles found in the coralline red algae (28). Layered structures were observed in both *O.wendtii* and *Jania* sp. and were related to alternating concentrations of high-Mg nanoparticles in the layers (27, 28).

In all of the above organisms studied, regardless of the biological and morphological differences between them, the presence of high-Mg calcite nanoparticles led to enhanced fracture resistance. The use of hierarchical structures to enhance mechanical properties is a well-known strategy exploited in nature by various organisms (39–43). Based on our earlier findings, we envisaged that the inclusion of high-Mg nanoparticles within Mg-calcite crystals formed via spinodal decomposition is a widespread strategy in nature, serving pivotal structural roles in enhancing the mechanical properties of the biomineralized tissue. In the present study we further generalize this biostrategy and confirm its widespread prevalence in high-Mg calcite biomineralization. We show that a widespread and relatively small brittle star (44) *Ophiactis savignyi*, as well as the soft coral *Tubipora musica* (also known as the organ pipe coral) and the starfish *Echinaster sepositus*, show evidence of high-Mg nanoparticles dispersed within a low-Mg calcite matrix. We carried out an extensive comparative study, which yielded fundamental insights into the similarities and differences in structure between the different organisms. We can now state that spinodal decomposition is indeed a general precrystallization step preceding the formation of high-Mg nanoparticles across different kingdoms and phyla, and that the specific chemical composition of the organism's skeletons is the basis for this phenomenon. The results of this study can be expected to profoundly affect the understanding of the strengthening mechanisms utilized by organisms to improve their mechanical endurance.

Results

This study was focused on investigating the structure of a variety of organisms across different kingdoms and phyla. The following organisms were selected: 2 coralline red algae species belonging to the plant kingdom, *Jania* sp. and *Corallina* sp.; and 7 organisms included in the animal kingdom but classified among different phyla, namely 2 brittle stars *Ophiomastix wendtii* and *Ophiactis savignyi*, 3 sea urchins *Phyllacanthus imperialis*, *Paracentrotus lividus* and *Heterocentrotus mammillatus* (all from the phylum Echinodermata), the soft coral *Tubipora musica*, also known as the organ pipe coral (phylum Cnidaria), and a starfish, *Echinaster sepositus* (phylum Echinodermata). Figure 1A−G displays the different organisms (insets) and their microstructures, imaged using high-resolution scanning electron microscopy (HRSEM). It is apparent from the images that the organisms are highly diverse. The brittle star *O. wendtii* (Figure 1A) has 5 arms and a body of approximately 1.5 cm, while *Jania* sp. and *Corallina* sp. (Figure 1B,C) are branched articulated algae comprised of calcified joints and interconnecting segments. The brittle star *O. savignyi* (Figure 1D) also possesses 5 arms but its body dimension of approximately 0.5 cm is smaller than that of *O. wendtii*. The skeleton of *T. musica* (Figure 1E) is arranged in hollow tubes with a diameter of approximately 0.2 cm and interconnecting horizontal platforms. The starfish *E. sepositus* (Figure 1F) possesses 5 arms. Microstructures of the mineralized tissues of the 6 organisms also differ significantly. Whereas *Jania* sp. and *Corallina* sp. possess nanometric-sized crystals deposited on the cell walls (see also supplementary materials, Figure S1), the crystalline features of the 4 other organisms are micrometric in size. The 2 coralline red algae have nanocrystals of similar dimensions. On the other hand, the crystalline features in the 2 brittle stars are different: the lenses of *O. wendtii* are typically 20−50 microns in size, whereas the features of *O. savignyi* are significantly smaller and are in the range of a few microns. Exceptional to all other organisms is *T. musica*, whose microstructure is significantly less porous.

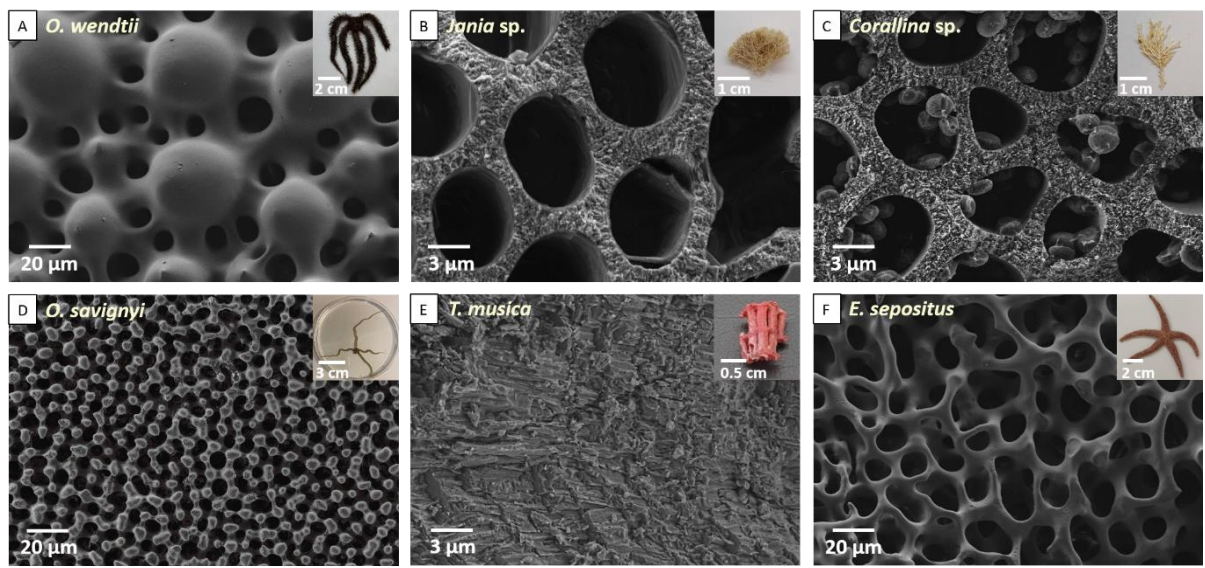

Figure 1- Microstructures of the different organisms imaged using HRSEM: A) *O. wendtii*. B) *Jania* sp. C) *Corallina* sp. D) *O. savignyi*. E) *T. musica*. F) *E. sepositus*. The different organisms are presented in the insets.

Mineralized tissues of each organism were studied in powdered form using synchrotron high-resolution powder X-ray diffraction (HRPXRD). The collected full diffraction patterns presented in Figure 2A indicated that all skeletons possess the crystalline structure of calcite. As represented by the {104} diffraction peak in Figure 2B, the diffraction peaks are shifted to higher 2θ angles relative to those of pure calcite. Based on our previous studies of other organisms, *O. wendtii* (9), *Jania* sp. (34) and *Corallina* sp. (28), this shift is probably also a result of substitution of Mg, with its smaller ionic radii, for Ca ions in the calcite lattice (24). Indeed, energy dispersive X-ray spectroscopy (EDS) analysis of the samples confirmed the presence of Mg in all the studied organisms (see supplementary materials, Figure S2). As expected owing to Mg substitution, lattice parameters extracted using the Rietveld refinement are smaller than those of pure calcite (see supplementary materials, Table S1). It is also possible that the shift of the diffraction peaks is a result of additional stresses in the crystals. A prominent difference is observed in the width of the {104} diffraction peak of the different organisms, with a broader peak in the case of *Jania* sp. and *Corallina* sp., which can be related to the nanometric size of their crystals (Figure 1B,C).

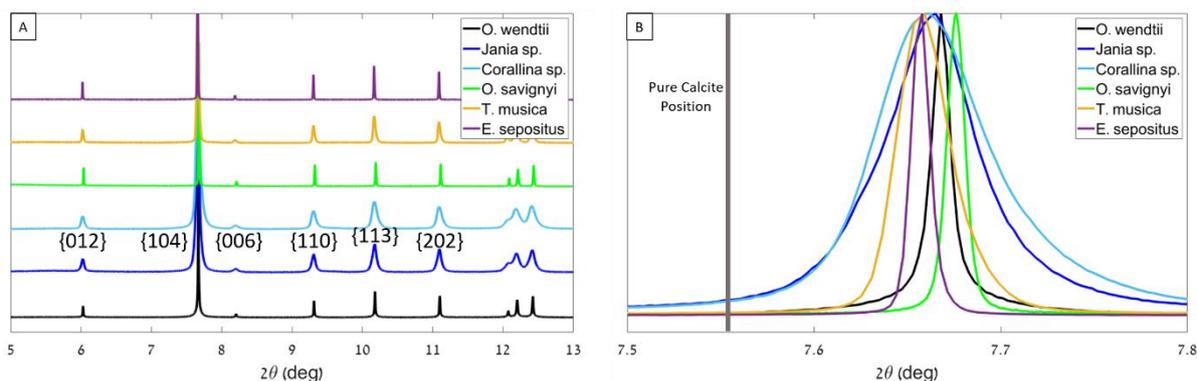

Figure 2- Synchrotron radiation HRPXRD data. A) HRPXRD patterns of the mineralized tissues of the studied organisms showing their crystalline calcite structures. The data were collected at a wavelength of 0.4 Å. B) Mg-calcite {104} diffraction peaks of the studied organisms relative to the {104} position of the pure calcite.

Previous studies by our group had revealed the presence of high-Mg nanoparticles dispersed within a low-Mg calcite matrix in the brittle star *O. wendtii* (9) and in the coralline red algae *Jania* sp. and *Corallina* sp. (28). Those findings were surprising, since the hosting Mg-calcite crystals had previously been considered single crystals, given the spot-like diffraction pattern observed under transmission electron microscopy (TEM). The high-Mg nanoparticles of *O. wendtii* were found in the previous work to be coherently aligned with the hosting Mg-calcite matrix and to exhibit a continuous lattice pattern in a phase-contrast high-resolution TEM (HRTEM) image (9). The high-Mg nanoparticles in *Jania* sp. and *Corallina* sp. were found, on the contrary, to be incoherent with the hosting matrices and to possess various crystallographic orientations in the phase-contrast HRTEM image. That conclusion had also been supported by the fact that fast Fourier transform (FFT) of the images acquired from a region containing both the matrix and the nanoparticles had demonstrated ring diffraction patterns (28). In our present quest to understand the origin of high-Mg nanoparticles in Mg-calcite systems and to generalize their presence as a widespread phenomenon in biomineralization, we have expanded our study and have shown, interestingly, that similar observations are found within other organisms. Within the low-Mg calcite matrices of the brittle star *O. savignyi*, the coral *T. musica* and the starfish *E. sepositus*, we detected high-Mg nanoparticles. All of the mineralized tissues diffracted as single crystals under TEM, yet examination of their nanostructures using HRTEM disclosed the presence of nanoparticles dispersed within their matrices (Figure 3). Figure 3A presents a HRTEM image of a lamella of the brittle star *O. savignyi* acquired using a high-angle annular dark field (HAADF) detector, and reveals nanoparticles displaying a darker contrast than that of the matrix. Owing to the sensitivity of the HAADF detector to changes in atomic number, this result indicates that the nanoparticles are composed of lighter elements than those of the hosting matrix. Combining this result with data from EDS again paints a picture of nanoparticles composed of Mg-calcite with a higher Mg content than that of the matrix. The nanoparticles, similarly to our previous findings on the high-Mg nanoparticles found in the brittle star *O. wendtii* (27), are arranged in layers. Although the presence of nanoparticles is evident within the matrix, the diffraction pattern of *O. savignyi* is that of a single crystal (Figure 3A, inset), similarly to our previous findings in *O. wendtii* and in *Jania* sp. and *Corallina* sp. Examination of a lamella of the coral *T. musica* using HRTEM presents the same phenomenon (Figure 3B). Crystalline nanoparticles are present within the Mg-calcite matrix and, according to the phase-contrast image, possess various crystallographic orientations, i.e., they are incoherent with the hosting matrix. This is even though a large area of the sample containing both the matrix and the nanoparticles diffracts as a single crystal (Figure 3B, inset). The starfish *E. sepositus* also exhibits nanoparticles, although the diffraction shows a spot diffraction pattern from a large area. The nanoparticles are coherent with the hosting matrix since both the matrix and the nanoparticles present a continuous pattern in phase-contrast HRTEM.

There is a clear difference in the arrangement of the nanoparticles found in the different studied organisms. The nanoparticles in the brittle star *O. wendtii* (9) and the starfish *E. sepositus* (present study) are coherent with the hosting matrix. The nanoparticles in *Jania* sp. (28), *Corallina* sp. (28) and *T. musica* (present study) are incoherent with the hosting matrix. We are not sure of the reason for this difference between species; we suspect, however, that it is governed by the kinetics of the spinodal decomposition and the subsequent nucleation and growth.

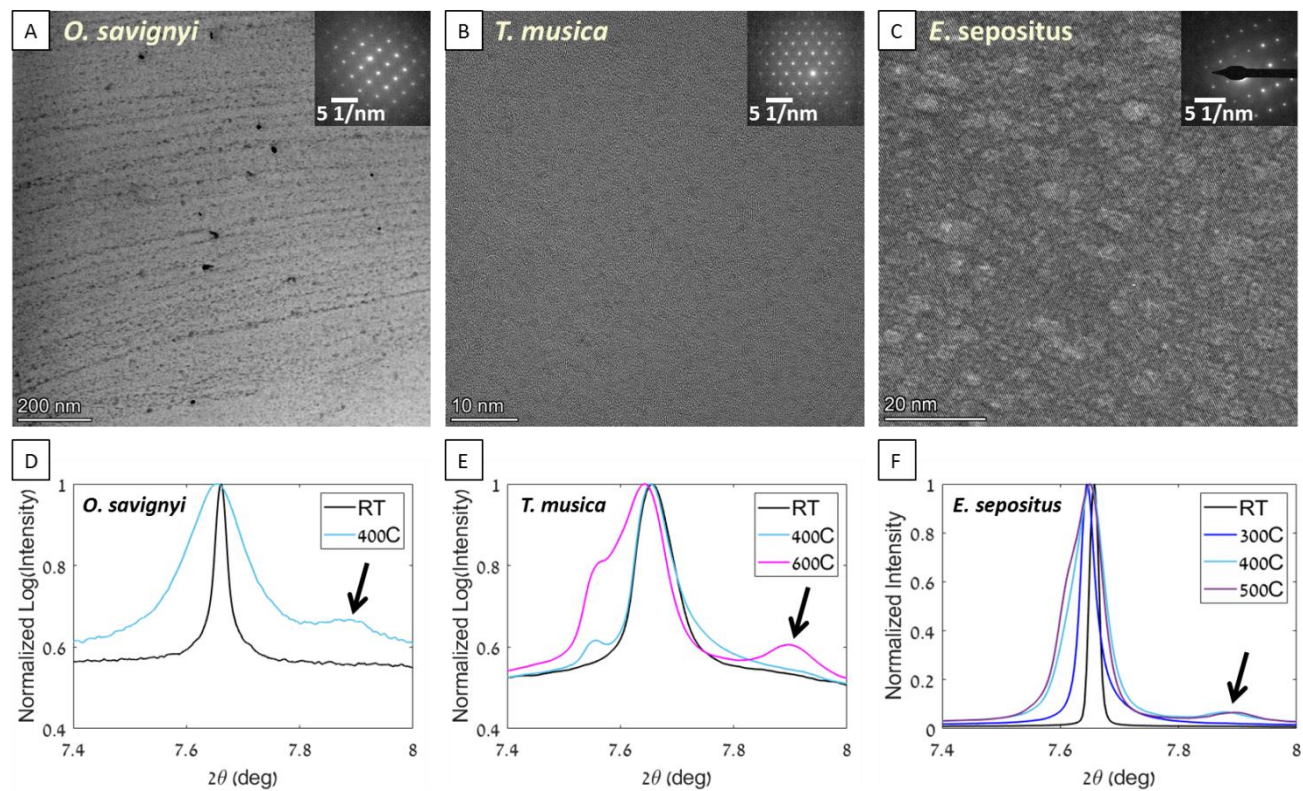

Figure 3- HRTEM and synchrotron radiation HRPXRD results. A) HRTEM image of *O. savignyi* acquired using a HAADF detector. B) HRTEM image of *T. musica* showing incoherent nanoparticles. C) HRTEM image of *E. sepositus* showing coherent nanoparticles. Insets: diffractions from regions containing the matrix as well as the nanoparticles in the studied organisms. HRPXRD results show the evolution of the {104} diffraction peak of Mg-calcite after heating, presented for a wavelength of 0.4 Å: D) *O. savignyi*. E) *T. musica*. F) *E. sepositus*.

The presence of nanoparticles within the lattice of all the organisms in the present study was also supported here by the results of isochronous annealing experiments coupled with synchrotron radiation HRPXRD. By means of this method the appearance of an additional broad diffraction peak after heating to 400°C had been established in *O. wendtii* for the first time (9). The 2-theta position of that diffraction peak corresponded to a {104} reflection of calcite with a relatively high Mg content of approximately 40 mol% (9), and its appearance was then attributed to the loss of coherence between the high-Mg nanoparticles and the matrix upon their heat-induced growth. Similarly (28), the appearance of a broad {104} diffraction peak at a higher 2-theta angle had been detected for *Corallina* sp. after heating to 600°C in a high-pressured $CO_2$ environment (approximately 30 atmospheres), although the same experiment conducted in a gaseous environment of air had resulted in no such outcome (28). In the high-pressured $CO_2$ environment, the Mg-calcite was able to stabilize, thus delaying its decomposition at this high temperature to Mg-calcite with a lower Mg content, MgO and $CO_2$. Given that the nanoparticles in *Corallina* sp. were found to be incoherent with the hosting matrix even at room temperature, the appearance of the broad diffraction peak at 600°C was attributed to coarsening of the nanoparticles at high temperatures and their larger average size after heating, as was indeed supported by an *in-situ* HRTEM heating experiment. The same experiments of isochronous annealing coupled with synchrotron radiation HRPXRD had been conducted in the case of *Jania* sp., both in air and in high-pressured $CO_2$, yet no additional diffraction peak had been detected, although the presence of nanoparticles had been clearly perceived on HRTEM. We, therefore, conjectured that the possible difference between the nanoparticles of the 2 coralline red algae might be explained in terms of their different thermal stabilities: *Jania* sp. nanoparticles were suggested to be less thermally stable, and hence not detected by the HRPXRD measurements (28).

Powdered samples of the organisms in the present study were heated to various temperatures and measured using synchrotron radiation HRPXRD at room temperature after the samples had cooled. The findings obtained in the new species were akin to the earlier observations in *O. wendtii*, *Corallina* sp. and *Jania* sp.. We found that after heating a sample of the brittle star *O. savignyi ex-situ* in air to 400°C, a new broad diffraction peak appeared at higher 2θ angles (Figure 3D) at a position corresponding to calcite with ~42 mol% Mg. The same was detected for the coral *T. musica,* but only after its heating *ex-situ* in air to the higher temperature of 600°C (Figure 3E). The Mg content related to the new diffraction peak was estimated as ~44 mol% Mg. To measure the starfish *E. sepositus* we used isochronous annealing coupled with synchrotron radiation HRPXRD in air (Figure 3F). The new diffraction peak appeared at a temperature of 400°C, which corresponds in position to a Mg content of ~41 mol% Mg.

Examination of all the studied organisms revealed some substantial differences in the thermal stabilities of the embedded high-Mg nanoparticles. In *O. wendtii*, *O. savignyi*, *T. musica* and *E. sepositus* the high-Mg nanoparticles appear as a distinct diffraction peak in HRPXRD after heating performed in an air environment, whereas those in *Corallina* sp. are apparent only when measured in high-pressured $CO_2$ of approximately 30 atmospheres and not in air (28). The stabilization of the Mg-calcite that occurs in high-pressured $CO_2$ allows stabilization of the nanoparticles and delay of their decomposition at high temperatures. In *Jania* sp., however, nanoparticles did not appear, either in air or in $CO_2$, as a distinct diffraction peak, thus indicating their decreased thermal stability comparably to that found within *Corallina* sp. (28). These results thus indicate a decreased thermal stability for the high-Mg nanoparticles found within the coralline red algae relative to the other studied organisms. A possible reason for this could be the different length-scales of the hosting matrix crystals, which are micrometric in *O. wendtii*, *O.* savignyi, T. *musica* and *E. sepositus*, whereas in the 2 studied coralline red algae they are nanometric. Their nanometric size in the red algae could possibly be attributable to a high nucleation rate, which might also result in a noncoherent arrangement of the nanoparticles.

The HRTEM and HRPXRD results confirmed that 6 of the 9 studied organisms include high-Mg nanoparticles within matrices composed of Mg-calcite with a lower Mg content, despite the fact that biologically they are highly diverse. This strongly suggests that the presence of high-Mg nanoparticles within Mg-calcite hosting crystals is a widespread phenomenon in nature. Although the presence of high-Mg nanoparticles has been established in many mineralized tissues, this was not the case for all organisms studied. Spines of the sea urchins *P. imperialis*, *P. lividus* and *H. mammillatus* did not demonstrate the appearance of an additional broad diffraction peak upon isochronous annealing coupled with synchrotron radiation HRPXRD, even in a high-pressured $CO_2$ environment (see supplementary materials, Figure S3).

The fact that high-Mg nanoparticles are selectively present within the different species and even within species of the same phylum (as shown here for the studied brittle stars compared to the studied sea urchins, all echinoderms) is a strong indication that this phenomenon is not related to biological similarities between species, but can rather be attributed to the compositional similarities of their mineralized tissues and specifically to the Mg content. To verify this supposition we analyzed all samples chemically, using inductively coupled plasma optical emission spectroscopy (ICP-OES) to measure the average Mg contents within the mineralized tissues. The results, presented in Figure 4, confirm the presence of Mg within all samples. The studied organisms can be separated into 2 categories. For all those in which high-Mg nanoparticles were detected (*O. wendtii*, *Jania* sp., *Corallina* sp., *O. savignyi*, *T. musica* and *E. sepositus*), the Mg content is relatively high and in the range of 15.20−16.42 mol% (calculated based on the ratio of Mg/(Ca+Mg)). For those organisms in which a new broad diffraction peak was not detected in the HRPXRD (the spines of *P. imperialis*, *P. lividus* and *H. mammillatus*), the Mg content was substantially lower and in the range of 4.11−10.63 mol%. As proposed in our previous study (27), the high-Mg nanoparticles in *O. wendtii* are formed via spinodal decomposition of the Mg-ACC precursor. Generally, spinodal decomposition was shown to be enabled when the Mg content in the calcite is in the range of 14 to 48 mol%, i.e., in high-Mg calcite systems. Given the high Mg contents within the calcitic skeletons (exceeding the thermodynamically stable limit of Mg within calcite) studied herein , and owing to the intricate microstructures found in these organisms, it is reasonable to suggest that their crystallization followed from a Mg-ACC precursor, as was previously stated for many biominerals, among them *O. wendtii, Jania* sp.*,* and *Corallina* sp. (9, 20, 28, 45–47). The present detection of high-Mg nanoparticles in *O. savignyi*, *T. musica* and *E. sepositus*, along with the fact that their skeletons all include Mg contents within a range that allows spinodal decomposition, are strong indications that the formation of high-Mg nanoparticles within Mg-calcite crystals follows spinodal decomposition of a Mg-ACC precursor, and is a widespread phenomenon that is not limited to specific organisms.

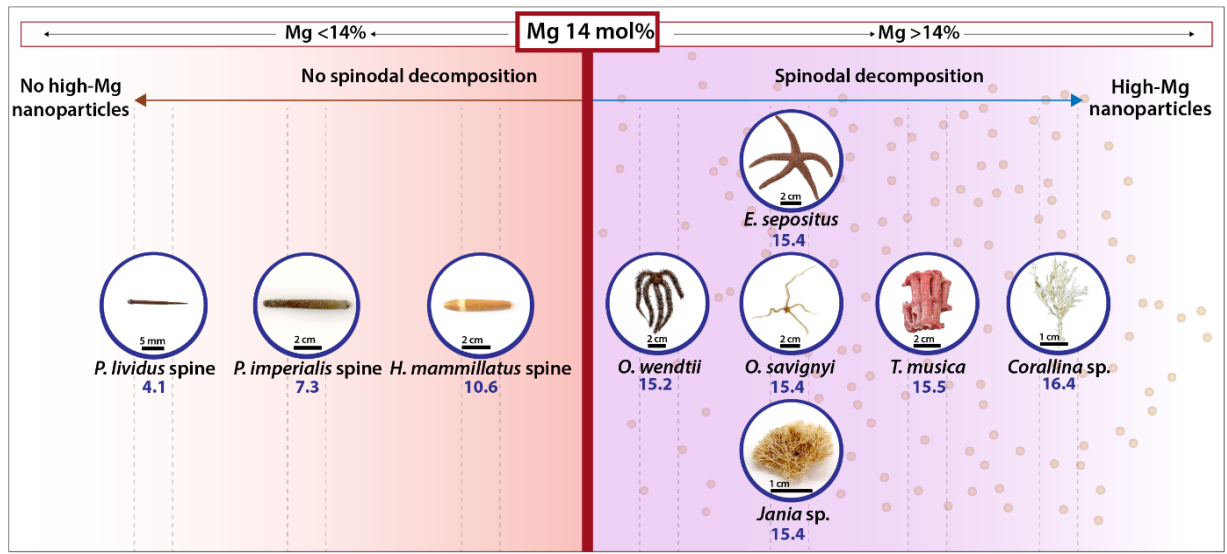

Figure 4. Inductively coupled plasma optical emission spectroscopy (ICP-OES) results presenting Mg contents in the studied organisms (defined as the ratio Mg/(Ca+Mg)). In this figure, 14 mol% Mg is defined—according to Seknazi et al. (27)—as the lowest limit required for spinodal decomposition.

Discussion

Based on the findings from our previous and current research on the structure of biomineralized tissues in the 9 selected organisms, we conclude that the formation of high-Mg calcite nanoparticles within a low-Mg calcite matrix via spinodal decomposition of a Mg-ACC precursor is a widespread phenomenon in high-Mg calcite biological systems. Contemplating the bigger picture, we propose that the formation of Mg-rich nanoparticles within a matrix of lower Mg content, as evident only in those organisms in which the Mg content exceeds the limit of 14 mol%, is already occurring via spinodal decomposition in the amorphous precursor. This is further supported by the fact that in those organisms in which the Mg concentration is lower than 14 mol% no such nanoparticles exist. Despite the great biological diversity of the studied organisms from different kingdoms and phyla, their common factor is their compositional similarity.

The high-Mg nanoparticles existing in the different organisms play critical functional roles in the biomineralization process and have implications for the mechanical properties of those organisms. They contribute to an increased fracture toughness and increased crack resistance of the organisms' skeletons, which in turn enables the organisms to withstand external stresses and possible dangers from their natural environment.

This unique and previously unknown strengthening strategy is found to be widespread in nature and is of great importance in the field of materials science, with special emphasis on the design of high-performance tough materials.

Materials and Methods

*Sample collection:* Samples of *O. wendtii* were collected in Belize. Samples of *Jania* sp. and *Corallina* sp. were collected from the shallow waters of the Mediterranean Sea, Israel. Samples of *O. savignyi* were collected from the Red Sea, Israel. Samples of *T. musica* were collected in Zanzibar. Samples of the starfish *E. sepositus* were collected from the Mediterranean Sea, Israel. Samples of *P. imperialis* and *H. mammillatus* were collected in the Philippines. Samples of *P. lividus* were supplied by the Israel Oceanographic and Limnological Research Institute.

*Sample bleaching:* Samples of *O. wendtii* and *E. sepositus* were bleached for the removal of organic matter in a deionized (DI) water and sodium hypochlorite (NaOCl) solution with a 2:1 volume ratio of NaOCl:DI for 8 h and 2 h, respectively. Sodium carbonate ($Na_2CO_3$, 2 wt%) was added to the DI water to avoid dissolution of calcite. After bleaching of the organics, the mineralized samples were washed several times with DI and dried in air.

*HRSEM:* Microstructures of the 6 studied organisms were examined by HRSEM using a Zeiss Ultra-Plus FEG-SEM, with a secondary electron (SE) detector and acceleration voltages of 4 kV. Prior to measurement the samples were coated with carbon.

*EDS:* EDS was performed using an Oxford Silicon Drift Detector EDS installed in a Zeiss Ultra-Plus FEG-SEM, with an acceleration voltage of 10 kV.

*HRPXRD:* Samples of *O. wendtii*, *Jania* sp. and *Corallina* sp. were measured by synchrotron radiation HRPXRD at the ID22 beamline of the European Synchrotron Radiation Facility (ESRF) in Grenoble, France, using a wavelength of 0.3999 Å at room temperature. Samples of *O. savignyi* were measured using synchrotron radiation HRPXRD at the 11BM-B beamline at the Advanced Photon Source (APS) at the U.S. Department of Energy's Argonne National Laboratory, Illinois, U.S.A., using a wavelength of 0.4579 Å. Heating was performed *ex-situ* at 400 °C for 30 min and the sample was later measured at room temperature after heating. Samples of *T. musica* were measured using synchrotron radiation HRPXRD at the ID22 beamline of the ESRF in Grenoble, France, using a wavelength of 0.3545 Å. Heating was performed *ex-situ* at 400 °C and 600 °C for 30 min and the samples were later measured at room temperature after heating. A sample of *E. sepositus* was measured using synchrotron radiation HRPXRD at the ID22 beamline of the ESRF in Grenoble, France, using a wavelength of 0.3542 Å. Isochronous annealing was performed to temperatures of 300 °C, 400 °C and 500 °C for 30 min. A sample of *P. imperialis* was measured using synchrotron radiation HRPXRD at the ID22 beamline of the ESRF in Grenoble, France, in a high-pressured $CO_2$ environment (approximately 20 atmospheres), using a wavelength of 0.3542 Å. Isochronous annealing was performed to temperatures of 400 °C, 500 °C and 600 °C for 30 min. A sample of *P. lividus* was measured using synchrotron radiation HRPXRD at the ID22 beamline of the ESRF in Grenoble, France, in a high-pressured $CO_2$ environment (approximately 15 atmospheres), using a wavelength of 0.3542 Å. Isochronous annealing was performed to temperatures of 400

°C, 500 °C and 600 °C for 30 min. A sample of *H. mammillatus* was measured using synchrotron radiation HRPXRD at the ID22 beamline of the ESRF in Grenoble, France, in a high-pressured $CO_2$ environment (approximately 30 atmospheres), using a wavelength of 0.3542 Å. Isochronous annealing was performed to temperatures of 500 °C and 600 °C for 30 min.

*HRTEM:* The FEI Titan Cubed Themis G2 60–300 was operated at 200 KV for the study of a FIB-sectioned lamella of *O. savignyi* in STEM mode using a HAADF detector, and at 60 KV for its study in TEM mode. The FEI Titan Cubed Themis G2 60–300 was operated at 60KV and 200 KV for the study of a plasma-FIB sectioned *T. musica* lamella in TEM mode. A FIB-sectioned lamella of *E. sepositus* was studied using the FEI Titan Cubed Themis G2 60–300 operated at 60 KV in TEM mode.

*ICP-OES:* ICP-OES was performed using an iCAP 6300 Duo ICP-OES spectrometer (Thermo Scientific). Prior to measurements the samples were weighed and fully dissolved in $HNO_3$.

Acknowledgments

The authors acknowledge the ID22 beamline at the ESRF, Grenoble, France, and the 11BM-B beamline at the APS at the U.S. Department of Energy's Argonne National Laboratory, Illinois, U.S.A. for assisting in collection high-resolution powder X-ray diffraction data. The authors are grateful to Dr. Gordon Hendler, Prof. Giuseppe Falini, Dr. Tali Mass and Dr. Boaz Mayzel for sample collection. This research was partly funded by the European Research Council under the European Union's Seventh Framework Program (FP/2013–2018)/ERC Grant Agreement No. 336077.

**Supporting Information**

High-Mg Calcite Nanoparticles Within a Low-Mg Calcite Matrix via Spinodal Decomposition: A Widespread Phenomenon

Nuphar Bianco-Stein[a], Iryna Polishchuk[a], Arad Lang[a], Lotan Portal[a], Catherine Dejoie[b] and Boaz Pokroy[a]*

[a] Department of Materials Science and Engineering and the Russell Berrie Nanotechnology Institute, Technion−Israel Institute of Technology, 32000 Haifa, Israel.

[b] ESRF-The European Synchrotron, CS 40220, 38043 Grenoble Cedex 9, France.

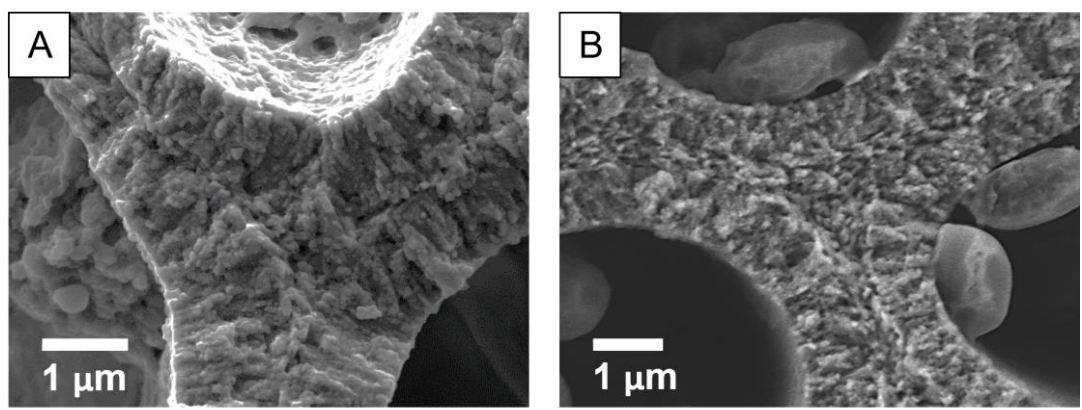

**Figure S1-** Nanometric-sized crystals deposited on the cell walls of A) *Jania* sp. and B) *Corallina* sp.

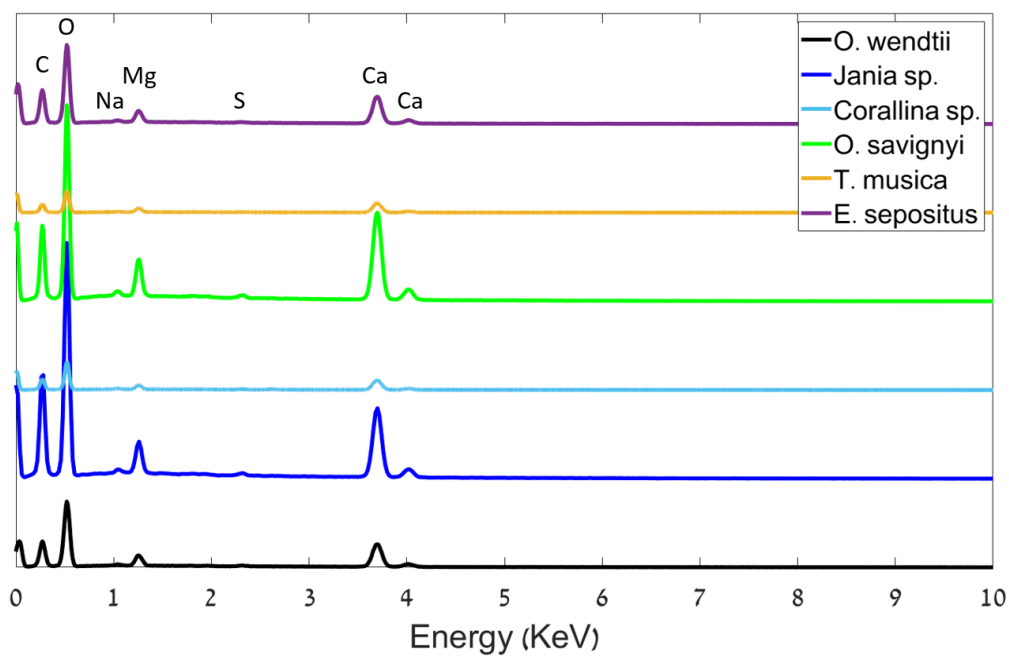

**Figure S2-** EDS spectra of the different organisms showing the presence of Ca and Mg.

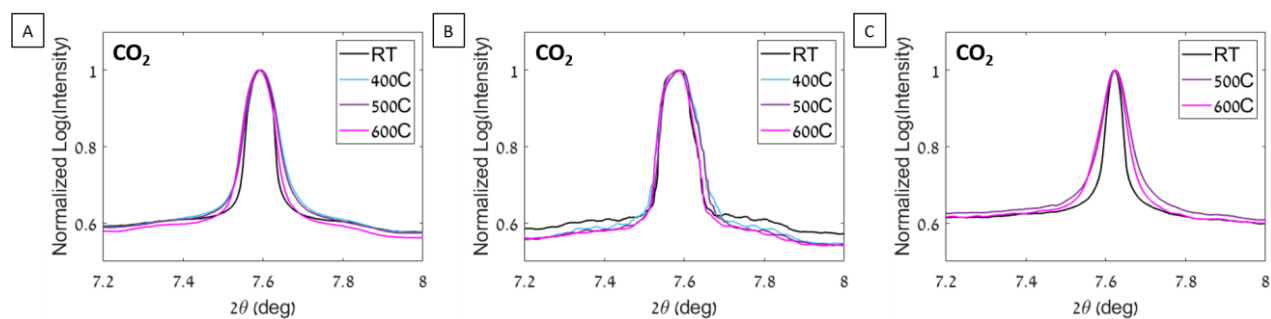

**Figure S3-** Spines of A) *Phyllacanthus imperialis* and B) *Paracentrotus lividus* C) Heterocentrotus mammillatus in $CO_2$ showing that there is no additional peak after heating. Shown at a wavelength of 0.4 Å. (1)

**Table S1-** Lattice parameters of the organisms calculated using the Rietveld refinement.

| Organism | *a*-lattice parameter | *c*-lattice parameter |
|---|---|---|
| Pure calcite(2) | 4.9903(4) | 17.0721(9) |
| *Ophiomastix wendtii* (3) | 4.92577(3) Å | 16.76897(6) Å |
| *Jania* sp. (4) | 4.9305(3) Å | 16.7904(5) Å |
| *Corallina* sp. | 4.92820(6) Å | 16.78472(5) Å |
| *Ophiactis savignyi* | 4.93092(3) Å | 16.79241(2) Å |
| *Tubipora musica* | 4.93166(4) Å | 16.80646(9) Å |
| *Echinaster sepositus* | 4.93245(0) Å | 16.80584(5) Å |